\documentclass[prd,onecolumn,showpacs,showkeys,amsmath,amssymb]{revtex4}

\newcommand{\vc}[1]{{\bf#1}}  
\newcommand{\ocn}[1]{{\mathcal{O}\left(c^{-#1}\right)}}
\begin{document}

\title{On the $1/c$ Expansion of $f(R)$ Gravity}

\author{Joachim N\"af} 
\email{naef@physik.uzh.ch}

\author{Philippe Jetzer}

\affiliation{Institut f\"{u}r Theoretische Physik, Universit\"{a}t Z\"{u}rich,
Winterthurerstrasse 190, CH-8057 Z\"{u}rich, Switzerland.}

\date{April 9, 2010}

\begin{abstract}
We derive for applications to isolated systems - on the scale of the 
Solar System - the first relativistic 
terms in the $1/c$ expansion of the space time metric $g_{\mu\nu}$ for 
metric $f(R)$ gravity theories, where $f$ is assumed to be analytic at $R=0$. For
our purpose it suffices to take into account up to quadratic terms
in the expansion of $f(R)$, thus we can approximate $f(R) = R + aR^2$ with a positive 
dimensional parameter $a$. 
In the non-relativistic limit, we get an additional Yukawa correction 
with coupling strength $G/3$ and Compton wave length $\sqrt{6a}$ to the Newtonian 
potential, which is a known result in the literature. 
As an application, we derive to the same order the correction to the geodetic 
precession of a gyroscope in a gravitational field and the precession of binary pulsars. 
The result of the Gravity Probe B experiment yields the limit  
$a \lesssim 5 \times 10^{11} \, \mathrm{m}^2$, whereas for the pulsar B 
in the PSR J0737-3039 system we get
a bound which is about $10^4$ times larger.
On the other hand the E\"ot-Wash 
experiment provides the best laboratory bound $a \lesssim 10^{-10} \, \mathrm{m}^2$.
Although the former bounds from geodesic precession are much larger than the laboratory 
ones, they are still meaningful in the case some type of chameleon effect is present 
and thus the effective values could be different at different length scales. 

\end{abstract}


\maketitle


\section{Introduction}
Since the emergence of the concepts of dark matter (DM) and dark energy (DE), 
they still lack in a concrete and satisfying physical model. This open question
motivated the development of new gravity theories. Most of 
them are direct modifications of general relativity (GR), which is still the 
simplest relativistic gravity theory fitting very accurately many precision
measurements in astrophysics, such as Mercury perihelion shift or mass diagrams of
double pulsars. Among such modified theories a lot of attention has been
devoted to the so-called metric
$f(R)$ theories with an action
\begin{equation}\label{act}
S = \frac{c^3}{16\pi G}\int f(R)\sqrt{-g}\,d^4x + S_M,
\end{equation}
where in contrast to GR the Einstein-Hilbert Lagrangian density is replaced by 
a nonlinear function $f(R)$. $S_M$ is the standard matter action. 
For an overview one may consult e. g. \cite{STR2} 
and references therein. 

In the literature there are several approaches which address the question of the 
non-relativistic limit as well as relativistic approximations of metric $f(R)$ theories. 
A discussion of the first relativistic corrections after a 
transformation to the Einstein frame is given in \cite{CAP1,CAP5}. 
The non-relativistic limit in the Jordan frame is investigated in 
\cite{CAP2,CAP3,CAP4}, whereas a calculation in the Palatini formalism is 
given for example in \cite{SOT}. 
In the present paper we work strictly in the Jordan frame. Our work is 
mainly motivated by the fact that the parametric post-Newtonian (PPN) formalism 
is not adapted to cover the $1/c$ expansion of $f(R)$ gravity \cite{CAPP}. 
As pointed out in \cite{CAP1,CAP2}, the corresponding non-relativistic limit 
indeed is not Newtonian, but contains a Yukawa type correction, too. We therefore 
derive the lowest order relativistic terms of the $1/c$-expansion of the space time 
metric governed by the Euler Lagrange equations of (\ref{act}). We thus achieve 
a ``post-Yukawa'' approximation of $f(R)$ gravity. This approximation 
is analogous to the complete first post-Newtonian approximation of GR, cf. 
for example \cite{WILL,WEIN,STR1,LL}. 

In section II we present the field equations of the model. Section III is 
devoted to the calculation of the expansion coefficients, and in section IV 
we make some remarks on the non-relativistic limit as well as the GR limit of 
the model. In section V we derive the equations of motion for a test 
particle and determine the underlying potentials for a set of freely falling particles. 
In section VI we apply our results to the precession of orbiting gyroscopes, 
and by using the measurements of Gravity Probe B and of the pulsar B in the 
PSR J0737-3039 system, we get upper limits for the value of $a$.

As far as notation is concerned: Greek letters denote space time indices and 
range from $0$ to $3$, whereas Latin letters denote space indices and range 
from $1$ to $3$. We take the sum over repeated indices within a term. By an 
index ``$,\mu$" we denote the partial differentiation w.r.t to $x^{\mu}$, 
except for $\mu = 0$, where it denotes the differentiation w.r.t the time 
coordinate $t$ rather than the coordinate $x^0 = ct$.


\section{The Field Equations}

Consider a 4-dimensional Lorentz manifold with metric 
$g_{\mu\nu}$ of signature $(-,+,+,+)$. We write $g = \det{g_{\mu\nu}}$ 
and denote the Ricci tensor of $g_{\mu\nu}$ by $R_{\mu\nu}$. 
The variation of the action (\ref{act}) w. r. t. the metric 
yields the Euler-Lagrange equations
\begin{equation}\label{eleqo}
f'(R)R_{\mu\nu} - \frac{1}{2}f(R)g_{\mu\nu} - 
\nabla_{\mu}\nabla_{\nu}f'(R) + g_{\mu\nu}\square_g f'(R) = 
\frac{8\pi G}{c^4}T_{\mu\nu},
\end{equation}
where $R = g^{\mu\nu}R_{\mu\nu}$, 
$T_{\mu\nu} = (-2c/\sqrt{-g})(\delta S_M/\delta g^{\mu\nu})$ is the energy 
momentum tensor, $c$ the vacuum speed of light, $G$ Newton's constant, 
$\nabla_{\mu}$ the covariant derivative for $g_{\mu\nu}$ and 
$\square_g = \nabla^{\mu}\nabla_{\mu}$. Taking the trace of (\ref{eleqo}) we obtain
\begin{equation}\label{eleqotr}
3\square_g f'(R) + f'(R)R - 2f(R) = \frac{8\pi G}{c^4}T,
\end{equation}
where $T$ is the trace of $T_{\mu\nu}$.
Motivated by the post-Newtonian approximation of GR we calculate the 
coefficients of the expansion of $g_{\mu\nu}$ in powers of $c^{-1}$:
\begin{eqnarray}\label{exp}
g_{00} & = & -1 + {}^{(2)}h_{00} + {}^{(4)}h_{00} + \ocn{6}, \\ 
g_{0i} & = & {}^{(3)}h_{0i} + \ocn{5}, \nonumber \\
g_{ij} & = & \delta_{ij} + {}^{(2)}h_{ij} + \ocn{4}, \nonumber
\end{eqnarray}
where ${}^{(n)}h_{\mu\nu}$ denotes a quantity of order $\mathcal{O}(c^{-n})$. 
The Ricci scalar is at least of order $\mathcal{O}(c^{-2})$.
Thus, if we assume the function $f$ to be analytic at $R = 0$ with $f'(0)=1$, 
it suffices to consider the expansion
\begin{equation}\label{fR}
f(R) = -2\Lambda + R + a R^2, \quad a \neq 0,
\end{equation}
in order to calculate the coefficients of $g_{\mu\nu}$ up to the orders 
indicated in (\ref{exp}), since higher powers of $R$ would only contribute 
to higher orders in the equations of the perturbation expansion. Moreover, 
since we adopt an expansion about a flat background space time in (\ref{exp}), we 
ignore a possible cosmological constant $\Lambda$ in what follows. The 
influence of a nonvanishing $\Lambda$ on the applications in section VI is 
discussed in \cite{sj1}. As we will see later, the parameter $a$ has to be 
positive for many reasons.

It is convenient to introduce 
the scalar field $\phi := f'(R)$. Since $f''(R) \neq 0$ holds for our choice of 
$f(R)$ in (\ref{fR}), we can invert $f'(R)$ in order to show that $f(R)$ gravity 
is equivalent to the Brans-Dicke theory with a non vanishing potential term 
and Brans-Dicke parameter $\omega_{BD} = 0$. We define the scalar field 
$\varphi$ by $\phi = 1 + 2a\varphi$, where we have chosen the asymptotic value 
such that a renormalization of the Newton's constant will be redundant (cf. the 
end of section IVA). Then the equations
(\ref{eleqo}) and (\ref{eleqotr}) are equivalent to
\begin{eqnarray}\label{eleq1}
R_{\mu\nu} & = &
\frac{1}{1+2a\varphi}\left(\frac{8\pi G}{c^4}\left(T_{\mu\nu}-
\frac{1}{3}g_{\mu\nu}T\right) +\frac{1}{6}g_{\mu\nu}\varphi + 
a\left(\frac{1}{2}g_{\mu\nu}\varphi^2 + 2\nabla_{\mu}
\nabla_{\mu}\varphi\right)\right)
\\ \label{eleq2}
\square_g \varphi& = & \frac{4\pi G}{3ac^4}T + \frac{1}{6a}\varphi.
\end{eqnarray}
The field $\varphi$ thus has the effective mass $\hbar/(c\sqrt{6a})$.
From (\ref{fR}) we infer that the dimensionless quantity $a R$ should be 
small compared to $1$. This fact reflects the concept of the chameleon effect 
\cite{KhWe}, which states the possibility that the Compton wave length 
$\lambda = \sqrt{6a}$ of the field $\varphi$ is smaller or larger in regions 
with higher or lower matter density, respectively. 
We understand our $1/c$ expansion to be valid in a local region which 
has an approximately constant mean matter density, in the sense that we assume 
the parameter $a$ to be constant on the length scale
characteristic for later applications, in particular the geodetic precession. 
On the other hand, $a$ may vary for applications which have different length scales.


\section{The Expansion Coefficients}

We introduce space time coordinates $(\vc{x},t)$, where bold 
face letters denote three dimensional vectors. The expansion coefficients are 
functions of these coordinates. Denote the 3-dimensional Nabla operator by 
$\nabla = (\partial_1,\partial_2,\partial_3)$. We find the following expansions for 
$R_{\mu\nu}$ and $\varphi$,
\begin{eqnarray}\label{}
R_{00} & = & \frac{1}{2}\Bigg(-\nabla^2\;{}^{(2)}h_{00} 
- \nabla^2\;{}^{(4)}h_{00} 
+ \frac{2}{c}{}^{(3)}h_{0i,0i} - \frac{1}{c^2}{}^{(2)}h_{ii,00} \\
& & + {}^{(2)}h_{ij}{}^{(2)}h_{00,ij} + 
{}^{(2)}h_{00,i}\left({}^{(2)}h_{ij,j} - 
\frac{1}{2}{}^{(2)}h_{jj,i} - \frac{1}{2}{}^{(2)}h_{00,i}\right)\Bigg) 
+ \ocn{6}, \nonumber \\
R_{0i} & = &  \frac{1}{2}\Bigg(-\nabla^2\;{}^{(3)}h_{0i} - 
\frac{1}{c}{}^{(2)}h_{jj,0i} + {}^{(3)}h_{j0,ij} 
+ \frac{1}{c}{}^{(2)}h_{ij,0j}\Bigg) + \ocn{5}, 
\nonumber \\ 
R_{ij} & = &  \frac{1}{2}\Bigg(-\nabla^2\;{}^{(2)}h_{ij} 
- {}^{(2)}h_{00,ij} - {}^{(2)}h_{kk,ij} 
+ {}^{(2)}h_{ik,kj} + {}^{(2)}h_{kj,ki} \Bigg) + \ocn{4}, \nonumber \\
\varphi & = & {}^{(2)}\varphi + {}^{(4)}\varphi + \ocn{6}, \nonumber
\end{eqnarray}
and for the energy momentum tensor,
\begin{eqnarray}\label{}
T^{00} & = & {}^{(-2)}T^{00} + {}^{(0)}T^{00} + \ocn{2}, \\ 
T^{0i} & = & {}^{(-1)}T^{0i} + \ocn{1}, \nonumber \\ 
T^{ij} & = & {}^{(0)}T^{ij} + \ocn{2}. \nonumber
\end{eqnarray}
Equation (\ref{eleq2}) then yields in leading order the Yukawa equation
\begin{equation}\label{R2}
\nabla^2\;{}^{(2)}\varphi - \alpha^2\;{}^{(2)}\varphi = 
-\frac{8\pi G\alpha^2}{c^4}{}^{(-2)}T^{00},
\end{equation}
where we have defined $\alpha^2 := 1/(6a)$ for a real $\alpha$. Equation
(\ref{R2}) has the solution
\begin{equation}\label{R2s}
{}^{(2)}\varphi(\vc{x},t) = \frac{1}{c^2}V(\vc{x},t)
\end{equation}
with the potential
\begin{equation}\label{V}
V(\vc{x},t) := \frac{2G\alpha^2}{c^2}\int\frac{{}^{(-2)}T^{00}(\vc{x}',t)
e^{-\alpha\left|\vc{x}-\vc{x}'\right|}}{\left|\vc{x}-\vc{x}'\right|}d^3x'.
\end{equation}
The $00$-component of equation (\ref{eleq1}) at order $\mathcal{O}(c^{-2})$ 
is given by
\begin{equation}\label{h002}
\nabla^2\;{}^{(2)}h_{00} = -\frac{32\pi G}{3c^4}{}^{(-2)}T^{00} 
+ \frac{1}{3}{}^{(2)}\varphi
\end{equation}
and has the solution
\begin{equation}\label{h002s}
{}^{(2)}h_{00}(\vc{x},t) = \frac{1}{c^2}\left(2U(\vc{x},t) - W(\vc{x},t)\right)
\end{equation}
with the potentials
\begin{eqnarray}\label{pot1}
U(\vc{x},t) & := & \frac{4G}{3c^2}\int\frac{{}^{(-2)}T^{00}(\vc{x}',t)}
{\left|\vc{x}-\vc{x}'\right|}d^3x', \\
W(\vc{x},t) & := & \frac{1}{12\pi}\int\frac{V(\vc{x}',t)}
{\left|\vc{x}-\vc{x}'\right|}d^3x'. \nonumber
\end{eqnarray}
In contrast to iterated Coulomb integrals, the potential $W$ is well defined 
because of the exponential decay of the Yukawa term in $V$. Before we determine the coefficients of 
the $0i$- and $ij$-components, we impose the four gauge conditions
\begin{eqnarray}\label{gauge1}
& & g_{ij,j} - \frac{1}{2}\left(g_{jj}-g_{00}\right)_{,i} 
- \frac{1}{3\alpha^2}\varphi_{,i} = \ocn{4}, \\
& & g_{0j,j} - \frac{1}{2c}g_{jj,0} - \frac{1}{3\alpha^2c}\varphi_{,0} \label{gauge2}
= \ocn{5}. 
\end{eqnarray}
Using the condition (\ref{gauge1}), we find for the $ij$-component of equation 
(\ref{eleq1}), up to order $\mathcal{O}(c^{-2})$,
\begin{equation}
\nabla^2\;{}^{(2)}h_{ij} = -\left(\frac{16\pi G}{3c^4}{}^{(-2)}T^{00} 
+ \frac{1}{3}{}^{(2)}\varphi\right)\delta_{ij}
\end{equation}
with the solution
\begin{equation}\label{hijs}
{}^{(2)}h_{ij}(\vc{x},t) = \frac{\delta_{ij}}{c^2}\left(U(\vc{x},t) + W(\vc{x},t)\right).
\end{equation}
Taking into account the gauge condition (\ref{gauge2}), the $0i$-component of equation 
(\ref{eleq1}), up to order $\mathcal{O}(c^{-3})$, simplifies to
\begin{equation}\label{h0i1}
\nabla^2\;{}^{(3)}h_{0i} = \frac{16\pi G}{3c^4}{}^{(-1)}T^{0i} 
+ \frac{1}{c}{}^{(2)}h_{ij,0j} - \frac{1}{2c}{}^{(2)}h_{jj,0i}.
\end{equation}
Defining the potentials
\begin{eqnarray}
\chi(\vc{x},t) & := & \frac{G}{3c^2}\int{}^{(-2)}T^{00}(\vc{x}',t)
\left|\vc{x}-\vc{x}'\right|d^3x', \\
\psi(\vc{x},t) & := & \frac{1}{48\pi}\int V(\vc{x}',t)\left|\vc{x}-\vc{x}'\right|d^3x',
\nonumber
\end{eqnarray}
we can write equation (\ref{h0i1}) as
\begin{equation}\label{h0i2}
\nabla^2\left({}^{(3)}h_{0i} + \frac{1}{2c^3}\left(\chi + \psi\right)_{,0i} \right) 
= \frac{16\pi G}{3c^4}{}^{(-2)}T^{0i}.
\end{equation}
The solution of equation (\ref{h0i2}) is given by
\begin{equation}
{}^{(3)}h_{0i}(\vc{x},t) = \frac{1}{c^3}\left(Y_i(\vc{x},t) + Z_i(\vc{x},t)\right)~,
\end{equation}
where 
\begin{eqnarray}
Y_i(\vc{x},t) & := & -\frac{4G}{c}\int\frac{{}^{(-1)}T^{0i}(\vc{x}',t)}
{\left|\vc{x}-\vc{x}'\right|}d^3x', \\
Z_i(\vc{x},t) & := & -\left(\chi(\vc{x},t) + \psi(\vc{x},t)\right)_{,0i}.
\nonumber
\end{eqnarray}
Thus, the $f(R)$ correction to the shift is only due to the gradient field 
$\vc{Z}$.

For the derivation of the component ${}^{(4)}h_{00}$, we first address
${}^{(4)}\varphi$. In view of the gauge (\ref{gauge1}) the 
$\mathcal{O}(c^{-4})$ part of equation (\ref{eleq2}) is given by
\begin{eqnarray}\label{R41}
\nabla^2\;{}^{(4)}\varphi - \alpha^2\;{}^{(4)}\varphi & = & 
\frac{8\pi G\alpha^2}{c^4}\left(-{}^{(0)}T^{00} + {}^{(0)}T^{ii}\right) \\
& & + \frac{1}{c^2}{}^{(2)}\varphi_{,00} + {}^{(2)}h_{ij}{}^{(2)}\varphi_{,ij} 
+ \frac{1}{6\alpha^2}{}^{(2)}\varphi_{,i}{}^{(2)}\varphi_{,i}. \nonumber
\end{eqnarray}
We use the identity
\begin{equation}\label{form}
\tilde{U}_{,i}\tilde{V}_{,i} = \frac{1}{2}\left(\nabla^2(\tilde{U}\tilde{V}) 
- \tilde{V}\nabla^2 \tilde{U} - \tilde{U}\nabla^2 \tilde{V}\right)
\end{equation}
for two arbitrary potentials $\tilde{U}$ and $\tilde{V}$ as well as the equations 
(\ref{R2}), (\ref{R2s}) and (\ref{hijs}) to rewrite equation (\ref{R41}) as
\begin{equation}\label{R42}
\nabla^2\left({}^{(4)}\varphi - \frac{1}{12\alpha^2c^4}V^2\right) 
- \alpha^2\left({}^{(4)}\varphi - \frac{1}{12\alpha^2c^4}V^2\right) = 
\frac{1}{c^4}A,
\end{equation}
where
\begin{eqnarray}
A & := & 8\pi G\alpha^2\left(-{}^{(0)}T^{00} + {}^{(0)}T^{ii} - 
\frac{1}{c^2}\left(U + W - \frac{1}{6\alpha^2}V\right){}^{(-2)}T^{00}\right) \\
& & + V_{,00}  + \alpha^2 UV + \alpha^2 VW - \frac{1}{12}V^2. \nonumber
\end{eqnarray}
From equation (\ref{R42}) we obtain
\begin{equation}\label{R4s}
{}^{(4)}\varphi(\vc{x},t) = \frac{1}{c^4}B(\vc{x},t),
\end{equation}
with the potential
\begin{equation}
B(\vc{x},t) = -\frac{1}{4\pi}\int\frac{A(\vc{x}',t)
e^{-\alpha\left|\vc{x}-\vc{x}'\right|}}{\left|\vc{x}-\vc{x}'\right|}d^3x'
+ \frac{1}{12\alpha^2}V^2(\vc{x},t).
\end{equation}
The $c^{-4}$-component of equation (\ref{eleq1}) can be simplified with 
equation (\ref{gauge2}):
\begin{eqnarray}\label{h0041}
\nabla^2\;{}^{(4)}h_{00} & = & {}^{(2)}h_{ij}{}^{(2)}h_{00,ij} + 
{}^{(2)}h_{00,i}\left({}^{(2)}h_{ij,j} - 
\frac{1}{2}{}^{(2)}h_{jj,i} - \frac{1}{2}{}^{(2)}h_{00,i}\right) \\
& & - \frac{16\pi G\alpha^2}{3c^4}\left(2{}^{(0)}T^{00} + {}^{(0)}T^{ii}
+ \left({}^{(2)}h_{00} - \frac{2}{3\alpha^2}{}^{(2)}\varphi\right){}^{(-2)}T^{00} 
\right) \nonumber \\
& & + \frac{1}{3}{}^{(4)}\varphi + \frac{1}{3}{}^{(2)}h_{00}{}^{(2)}\varphi
+ \frac{1}{18\alpha^2}\left({}^{(2)}\varphi\right)^2
-\frac{1}{3\alpha^2}{}^{(2)}h_{00,i}{}^{(2)}\varphi_{,i}. \nonumber
\end{eqnarray}
Using equations (\ref{R2}), (\ref{R2s}), (\ref{h002}), (\ref{h002s}),
(\ref{pot1}), (\ref{hijs}), (\ref{form}) and (\ref{R4s}) we write equation
(\ref{h0041}) in the form
\begin{equation}\label{h004}
\nabla^2\left({}^{(4)}h_{00} + \frac{1}{c^4}\left(\frac{3}{2}U^2
+ \frac{1}{3\alpha^2}UV - \frac{3}{4}UW - \frac{1}{6\alpha^2}VW\right)\right) =
\frac{1}{c^4}C
\end{equation}
with
\begin{eqnarray}
C & := & -\frac{16\pi G}{3}\left(2{}^{(0)}T^{00} + {}^{(0)}T^{ii}
+ \frac{1}{2c^2}\left(15U-\frac{2}{3\alpha^2}V\right){}^{(-2)}T^{00}\right) \\
& & + \frac{19}{12}UV - \frac{1}{6}VW + \frac{1}{9\alpha^2}V^2 + \frac{1}{3}B.
\nonumber
\end{eqnarray}
Hence,
\begin{equation}\label{h004s}
{}^{(4)}h_{00}(\vc{x},t) = \frac{1}{c^4}D(\vc{x},t),
\end{equation}
where
\begin{eqnarray}
D(\vc{x},t) & = & -\frac{1}{4\pi}\int\frac{C(\vc{x}',t)
}{\left|\vc{x}-\vc{x}'\right|}d^3x' - \frac{3}{2}U^2(\vc{x},t)
- \frac{1}{3\alpha^2}U(\vc{x},t)V(\vc{x},t) \\
& & + \frac{3}{4}U(\vc{x},t)W(\vc{x},t) +
\frac{1}{6\alpha^2}V(\vc{x},t)W(\vc{x},t). \nonumber
\end{eqnarray}
The metric field, up to the first relativistic approximation, is 
thus determined by the fields $U,W,\vc{Y},\vc{Z},D$.


\section{The GR Limit and the Non-Relativistic Limit}

\subsection{The GR Limit}

By taking the limit $a \to 0$ resp. $\alpha \to \infty$, the theory converges 
to GR. Notice that with the assumption $a > 0$ we have the following 
representation of the Dirac delta function:
\begin{equation}
\lim_{\alpha \to \infty}\frac{\alpha^2}{4\pi}\int\xi(\vc{x}')
\frac{e^{-\alpha\left|\vc{x}-\vc{x}'\right|}}
{\left|\vc{x}-\vc{x}'\right|}d^3x' = \int\xi(\vc{x}')
\delta(\vc{x} - \vc{x}')d^3x' = \xi(\vc{x})
\end{equation}
for an arbitrary test function $\xi(\vc{x})$. Hence
\begin{equation}
\lim_{\alpha \to\infty}W(\vc{x},t) =
\frac{2G}{3c^2}\int\frac{{}^{(-2)}T^{00}(\vc{x}',t)}
{\left|\vc{x}-\vc{x}'\right|}d^3x' = \frac{1}{2}U(\vc{x},t)
\end{equation}
and thus
\begin{eqnarray}
\lim_{\alpha \to \infty}{}^{(2)}h_{00}(\vc{x},t)  & = &
\lim_{\alpha \to \infty}{}^{(2)}h_{11}(\vc{x},t) \\
& = & \frac{3}{2c^2}U(\vc{x},t) =
\frac{2G}{c^4}\int\frac{{}^{(-2)}T^{00}(\vc{x}',t)}
{\left|\vc{x}-\vc{x}'\right|}d^3x' =\frac{2}{c^2}U_N(\vc{x},t), \nonumber
\end{eqnarray}
as expected. $U_N$ is exactly the Newtonian potential. There is no need for a 
rescaling of Newton's constant $G$, as it is sometimes the case for other 
modified gravity theories.

\subsection{The Non-Relativistic Limit}

From the equations (\ref{V}), (\ref{h002s}) and (\ref{pot1}) we gather that for the
$f(R)$ model given by (\ref{fR}) the non-relativistic limit is not Newtonian, 
since the component ${}^{(2)}h_{00}$ contains a Yukawa type term.

More explicitly, if for instance we consider a perfect, non viscous fluid with 
mass density $\rho$, pressure $p$ and velocity field $\vc{v} = (v^1,v^2,v^3)$, 
we have
\begin{eqnarray}\label{}
{}^{(-2)}T^{00} & = & \rho c^2, \\ 
{}^{(-1)}T^{0i} & = & \rho c v^i, \nonumber \\ 
{}^{(0)}T^{ij} & = & \rho v^i v^j + p\delta_{ij}. \nonumber
\end{eqnarray}
In the non-relativistic limit, the energy-momentum conservation
\begin{equation}\label{emc}
T^{\mu\nu}_{\phantom{\mu}\phantom{\nu};\nu} = 0
\end{equation}
then yields the equation of continuity,
\begin{equation}\label{cont}
\partial_t\rho + \partial_i(\rho v^i) = 0,
\end{equation}
and the analogous of the Euler equation,
\begin{equation}\label{eul}
\rho\left(\partial_t v^i + v^j\partial_j v^i\right) = -\partial_i p + 
\rho \partial_i\left(U - \frac{1}{2}W\right).
\end{equation}
Aside from a Newtonian term, the potential $W$ contains also a Yukawa type 
term. The gravitational force, represented by the second term on 
the right hand side, therefore contains also the gradient of a Yukawa 
potential. Together with the Poisson equations
\begin{eqnarray}\label{pois}
\nabla^2 U & = & -\frac{16\pi G}{3}\rho, \\
\nabla^2 W & = & -\frac{2G\alpha^2}{3}\int\frac{\rho(\vc{x}',t)
e^{-\alpha\left|\vc{x}-\vc{x}'\right|}}{\left|\vc{x}-\vc{x}'\right|}d^3x', 
\nonumber
\end{eqnarray}
(\ref{cont}) and (\ref{eul}) are the basic equations of the non-relativistic limit 
of hydrodynamics for metric $f(R)$ theory. The 
parameter $a = 1/(6\alpha^2)$ can be constrained by experiments which test a 
Yukawa type correction to the Newtonian potential. Experimental data and overviews 
can be found for example in \cite{EotWash,NEBEBO,GSW,sj2,KoMe,IOR}. Constraints 
on $f(R)$ theories are given for instance in \cite{BvDS}. From equation 
(\ref{eul}) (see also equation (\ref{pgpp})) we find for our specific 
model the Yukawa field strength $G/3$. The E\"ot-Wash experiment 
\cite{EotWash} thus yields the limit $a \lesssim 10^{-10} \, \mathrm{m}^2$.


\section{Particle Dynamics}

We can derive the equations of motion for a freely falling test particle 
in a field $(U,W,\vc{Y},\vc{Z},D)$ by evaluating the geodesic equation,
\begin{equation}\label{}
\frac{d^2x^{\mu}}{d\tau^2} + \Gamma^\mu_{\nu\lambda}
\frac{dx^{\nu}}{d\tau}\frac{dx^{\lambda}}{d\tau} = 0,
\end{equation}
where the $\Gamma^{\lambda}_{\mu\nu}$ denote the connection coefficients of the 
metric $g_{\mu\nu}$, and $x^{\mu}(\tau) = (ct(\tau),\vc{x}(\tau))$ is the 
position of the test particle at proper time $\tau$. Defining 
$\vc{v} := d\vc{x}/dt$, the equations of motion in vector notation read
\begin{eqnarray}\label{eqom}
\frac{d\vc{v}}{dt} & = & \nabla\left(U - \frac{1}{2}W\right) + 
\frac{1}{c^2}\Bigg((U + W)\nabla\left(U - \frac{1}{2}W\right) -\nabla D  \\
& & + \partial_t (\vc{Y} + \vc{Z}) + \vc{v} \wedge (\vc{\nabla} \wedge \vc{Y})
- \vc{v}\partial_t \left(2U +\frac{1}{2}W\right) \nonumber \\
& & - 3\vc{v}(\vc{v}\cdot\vc{\nabla})U 
- \vc{v}^2\nabla(U + W)\Bigg) + \ocn{4}. \nonumber
\end{eqnarray}

The energy momentum tensor of a set of point particles reads \cite{WEIN}
\begin{equation}
T^{\mu\nu}(\vc{x},t) = \frac{1}{\sqrt{-g}}\sum_n
m_n\frac{dx_n^{\mu}}{dt}\frac{dx_n^{\nu}}{dt}\left(\frac{d\tau_n
}{dt}\right)^{-1}\delta^3
\left(\vc{x}-\vc{x}_n(t)\right),
\end{equation}
where $m_n$ is the mass of the $n$-th particle, $\tau_n$ is its proper time 
and $\vc{x}_n(t) = (x_n^1(t),x_n^2(t),x_n^3(t))$ 
is its position at time $t$. This leads to the expansion
\begin{eqnarray}\label{empp}
T^{00} & = & \sum_n
m_n\left(c^2 + \frac{1}{2}\left(U - 5W + \vc{v}_n^2\right)\right)\delta^3
\left(\vc{x}-\vc{x}_n\right) + \ocn{2}, \\ 
T^{0i} & = & \sum_n m_n c \, v_n^i\delta^3
\left(\vc{x}-\vc{x}_n\right) + \ocn{1}, \nonumber \\ 
T^{ij} & = & \sum_n m_n v_n^i v_n^j\delta^3
\left(\vc{x}-\vc{x}_n\right) + \ocn{2}, \nonumber
\end{eqnarray}
where $\vc{v}_n = d\vc{x}_n/dt$. Inserting (\ref{empp}) into the defining equations 
of the potentials $U$, $W$, $\chi$, $\psi$ and $Y_i$, we get
\begin{eqnarray}\label{pppot}
U(\vc{x},t) & = & \frac{4G}{3}\sum_n \frac{m_n}
{\left|\vc{x}-\vc{x}_n(t)\right|}, \\
W(\vc{x},t) & = & \frac{2G}{3}\sum_n \frac{m_n}{\left|\vc{x}-\vc{x}_n(t)\right|}
\left(1 - e^{-\alpha\left|\vc{x}-\vc{x}_n(t)\right|}\right), \nonumber \\
\chi(\vc{x},t) & = & \frac{G}{3}\sum_n m_n
\left|\vc{x}-\vc{x}_n(t)\right|, \nonumber \\
\psi(\vc{x},t) & = & \frac{G}{6}\sum_n m_n\left(\left|\vc{x}-\vc{x}_n(t)\right| + 
\frac{2}{\alpha^2\left|\vc{x}-\vc{x}_n(t)\right|}
\left(1 - e^{-\alpha\left|\vc{x}-\vc{x}_n(t)\right|}\right)\right), \nonumber \\
Y_i(\vc{x},t) & = & -4G\sum_n \frac{m_n v_n^i(t)}
{\left|\vc{x}-\vc{x}_n(t)\right|}. \nonumber
\end{eqnarray}
Using (\ref{empp}) and (\ref{pppot}), one is able to calculate (at least formally) 
the potentials $\vc{Z}$ and $D$, which we however do not need for the applications 
we consider in this paper.

We investigate the non-relativistic limit by taking the $\mathcal{O}(1)$ part of 
equation (\ref{eqom}). Considering each particle as a test particle in the field 
of the other ones, we replace the potentials $U$ and $W$ by their self-energy 
free parts. The equations of motion for the test particle then read
\begin{equation}\label{pgpp}
\frac{dv_n^i}{dt} = G \sum_{k \neq n}\frac{\partial}{\partial x_n^i}\left(\frac{m_k}
{\left|\vc{x}_n - \vc{x}_k\right|}
\left(1 + \frac{1}{3}e^{-\alpha\left|\vc{x}_n - \vc{x}_k\right|}\right)\right).
\end{equation}
This is the analogue of the Newtonian equations of motion for a purely gravitating 
set of point particles.


\section{Precession of Orbiting Gyroscopes}

The following derivation of the precession and its applications is done in 
complete analogy with the corresponding computations in GR \cite{STR1,WEIN}. 
A recent and detailed review of gravitomagnetism in physics and astrophysics 
is provided in \cite{Scha}.

Consider a gyroscope with spin $\vc{S} = (S_1,S_2,S_3)$. We define its 3-velocity 
$\vc{v}$ to be the velocity of the non-relativistic center of mass of the gyroscope, 
the trajectory of which is assumed to be the one of a purely gravitating point particle. 
Notice that this last assumption includes the fact that the gyroscope moves along a 
geodesic and thus the Thomas precession due to a external force vanishes. The 
spin 4-vector $S_\mu := (S_0,\vc{S})$ precesses according to the 
equation of parallel transport,
\begin{equation}\label{pt1}
\frac{d S_\mu}{d\tau} = \Gamma^{\lambda}_{\mu\nu}S_\lambda\frac{dx^{\nu}}{d\tau},
\end{equation}
and satisfies the orthogonality condition 
\begin{equation}\label{orth}
\frac{dx^{\mu}}{d\tau} S_\mu = 0.
\end{equation}
Up to the lowest order, equation (\ref{pt1}) reads
\begin{equation}\label{pt2}
\frac{dS_i}{dt} = \left(c{}^{(3)}\Gamma^j_{i0} - {}^{(2)}\Gamma^0_{i0}v_j + 
{}^{(2)}\Gamma^j_{ik}v_k\right)S_j.
\end{equation}
Similarly as in GR we define the intrinsic spin vector by
\begin{equation}
\mathcal{S} := \left(1 - \frac{1}{2c^2}\left(U + W\right)\right) \vc{S} - 
\frac{1}{2c^2}\vc{v}(\vc{v}\cdot\vc{S}).
\end{equation}
Then $\mathcal{S}^2$ is an integral of (\ref{pt2}) up to the required order. 
Equations (\ref{pt2}) and (\ref{eqom}) then yield
\begin{equation}
\frac{d\mathcal{S}}{dt} = \vc{\Omega} \wedge \mathcal{S},
\end{equation}
and the precession angular velocity is given by 
\begin{equation}\label{omega}
\vc{\Omega} = -\frac{1}{2c^2}\left(\vc{\nabla} \wedge \vc{Y}\right) + 
\frac{1}{c^2}\vc{v} \wedge \vc{\nabla}\left(U + \frac{1}{4}W\right).
\end{equation}
Compared to GR, the Lense-Thirring precession represented by the first 
term remains unchanged. This was expected since, as mentioned 
in section III, a finite parameter $\alpha$ affects 
the shift only by a gradient field. On the other hand, 
the geodetic precession given by the second term is modified.

\subsection{Gyroscope Orbiting Around the Earth}

We now analyse the correction to the geodetic precession, since the 
Lense-Thirring precession is not modified.
We model the Earth as a sphere with mass $M$ which is at rest centred at 
the origin of our coordinate system. Consider the gyroscope to be in a 
circular orbit $\vc{x}(t)$ with radius $|\vc{x}(t)| \equiv r$ and unit 
normal $\vc{n}$, such that $(\vc{x},\vc{v},\vc{n})$ is a positively 
oriented dreibein. Then (\ref{pppot}) gives
\begin{equation}
U + \frac{1}{4}W = \frac{3GM}{2r}\left(1 - \frac{1}{9}e^{-\alpha r}\right),
\end{equation}
and by equating the gravitational and centrifugal force on the gyroscope, 
we find for the velocity
\begin{equation}
\vc{v} = \left(\frac{GM}{r^3}\left(1 + 
\frac{1}{3}\left(1 + \alpha r\right)e^{-\alpha r}\right)\right)^{1/2}\vc{n} 
\wedge \vc{x}.
\end{equation}
Hence the geodetic precession angular velocity is
\begin{eqnarray}\label{gpbres}
\vc{\Omega}_{\mathrm{geodesic}} & := & \frac{1}{c^2}\vc{v} \wedge 
\vc{\nabla}\left(U + \frac{1}{4}W\right) \\
& = & \frac{3(GM)^{3/2}}{2c^2 r^{5/2}}\left(1 + 
\frac{1}{3}\left(1 + \alpha r\right)e^{-\alpha r}\right)^{1/2}
\left(1 - \frac{1}{9}\left(1 + \alpha r\right)e^{-\alpha r}\right)\vc{n}. 
\nonumber
\end{eqnarray}
Obviously, $\vc{\Omega}_{\mathrm{geodesic}}$ converges to its GR value for 
$\alpha \to \infty$. This result can be compared with the 
measurements of the Gravity Probe B experiment \cite{GPB}. The measured value 
lies within a minimal residue of $30\;\mathrm{mas/yr}$ from the predicted GR 
value $6606\;\mathrm{mas/yr}$. This allows to constrain the relative 
deviation from the GR value in (\ref{gpbres}) by approximately $0.45\,\%$. 
Since this deviation decays faster than $e^{-\alpha r}$ with growing $\alpha$, 
while $r \approx 7 \times 10^6 \, \mathrm{m}$, we expect a much larger 
bound for $a$ as the one given by the E\"ot-Wash experiment. 
For the given accuracy of measurement, equation (\ref{gpbres}) yields 
$a \lesssim 5 \times 10^{11} \, \mathrm{m}^2$. 
If we estimate an upper limit for the scalar curvature using the mean 
Earth mass density, we are left with $R \lesssim 10^{-22}\,\mathrm{m}^{-2}$. 
Even in this very rough approximation we have $aR \ll 1$ for our constraint 
on $a$.

We remark that for a nonvanishing cosmological 
constant the residue in the Gravity Probe B measurements lead to
the bound $\Lambda \lesssim 3\times 10^{-27}\,\mathrm{m}^{-2}$
\cite{sj1}. When considering both $\Lambda \neq 0$ and the Yukawa term
within perturbation theory, to leading order the corrections due to $a$ and 
to $\Lambda$ will add linearly.


\subsection{Precession of Binary Pulsars}

Consider a binary system with center of mass at the origin. We index the mass 
$m_n$ and the position $\vc{x}_n$ by $1$ for the pulsar and by $2$ 
for the companion. Equation (\ref{pppot}) gives for the fields of the 
companion
\begin{eqnarray}
U(\vc{x},t) + \frac{1}{4}W(\vc{x},t) & = & \frac{3Gm_2}{2\left|\vc{x}-\vc{x}_2(t)\right|}
\left(1 - \frac{1}{9}e^{-\alpha \left|\vc{x}-\vc{x}_2(t)\right|}\right), \\
\vc{Y}(\vc{x},t) & = & -\frac{4Gm_2\vc{v}_2(t)}{\left|\vc{x}-\vc{x}_2(t)\right|}.
\end{eqnarray}
Define now $\vc{x} := \vc{x}_1 - \vc{x}_2$, $r := |\vc{x}|$, 
the reduced mass $\mu := m_1m_2/(m_1+m_2)$ and the 
angular momentum $\vc{L} := \mu\vc{x}\wedge(d\vc{x}/dt)$. Evaluating (\ref{omega}) 
at $\vc{x}_1$ then leads to
\begin{eqnarray}
\vc{\Omega} = -\frac{G\vc{L}}{c^2r^3}\left(2 + 
\frac{3m_2}{2m_1}\left(1 - \frac{1}{9}
\left(1 + \alpha r\right)e^{-\alpha r}\right)\right).
\end{eqnarray}
In order to approximate the average of $\vc{\Omega}$ over a period,
$\left<\vc{\Omega}\right> = T^{-1}\int_0^T\vc{\Omega}(t)dt$,
we need an expression of the trajectory in the non-relativistic limit. Therefore we 
deduce the correction to the Kepler ellipse
\begin{equation}\label{kepl}
r_0(\theta) := \frac{p}{1 + e\cos{\theta}},
\end{equation}
due to a small perturbation $\delta U$ of the 
Newtonian potential $U_N$. In (\ref{kepl}) we have introduced polar coordinates
$(r,\theta)$ and the parameters $p := \vc{L}^2/(Gm_1m_2\mu)$ and 
$e := (1 + 2Ep/(Gm_1m_2))^{1/2}$, where $E$ is the total energy of the two body 
system. 
We expand the equation of motion with respect to $\delta U$,
\begin{eqnarray}
\frac{dr}{d\theta} & = & 
\frac{r^2}{|\vc{L}|}\left(2\mu\left(E + 
\frac{Gm_1m_2}{r} + \delta U\right) - 
\frac{\vc{L}^2}{r^2}\right)^{1/2} \\
& = & \frac{r^2}{|\vc{L}|}\left(\left(2\mu\left(E + 
\frac{Gm_1m_2}{r}\right) - 
\frac{\vc{L}^2}{r^2}\right)^{1/2} + 
\mu\left(2\mu\left(E + 
\frac{Gm_1m_2}{r}\right) - 
\frac{\vc{L}^2}{r^2}\right)^{-1/2}\delta U\right) \nonumber \\
& & + \mathcal{O}(\delta U^2). \nonumber
\end{eqnarray}
Integration along the unperturbed trajectory $r_0(\theta)$ for 
$\delta U =  Gm_1m_2e^{-\alpha r}/(3r)$ yields 
$r(\theta) \approx r_0(\theta) + \delta r(\theta)$ with
\begin{equation}
\delta r(\theta) = \frac{Gm_1m_2\mu}{3|\vc{L}|}\int_0^\theta 
r_0(\tilde{\theta})e^{-\alpha r_0(\tilde{\theta})}
\left(2\mu\left(E  + \frac{Gm_1m_2}{r_0(\tilde{\theta})}\right) - 
\frac{\vc{L}^2}{r^2_0(\tilde{\theta})}\right)^{-1/2}d\tilde{\theta}.
\end{equation}
We approximate the average of $\vc{\Omega}$ as
\begin{eqnarray}\label{aver}
\left<\vc{\Omega}\right> & \approx & 
\frac{G\mu\hat{\vc{L}}}{c^2T}\int_0^{2\pi} 
\frac{1}{r}\left(2 + \frac{3m_2}{2m_1}\left(1 - 
\frac{1}{9}\left(1 + 
\alpha r_0\right)e^{-\alpha r_0}\right)\right)
d\theta  \\
& \approx & 
\frac{G\mu\hat{\vc{L}}}{c^2T}\left(\frac{2\pi}{p}\left(2 + \frac{3m_2}{2m_1}\right)
-\int_0^{2\pi}\left(\frac{\delta r}{r^2_0}\left(2 + 
\frac{3m_2}{2m_1}\right) + 
\frac{m_2}{6m_1r_0}\left(1 + 
\alpha r_0\right)e^{-\alpha r_0}\right)
d\theta\right). \nonumber
\end{eqnarray}
Due to the exponential decay the correction terms for $r$ and 
$\left<\vc{\Omega}\right>$ are very sensitive to 
variations of the parameter $a$. To give an idea of the orders of 
magnitude, we analyse (\ref{aver}) for the 
PSR 1913+16 data given in \cite{WeTa}. The predicted GR value given 
by the first term on the r.h.s of (\ref{aver}) evaluates to 
$1.21^{\circ}/\mathrm{yr}$. The correction term reaches approximately 
$1\%$ of the GR value for $a \approx 2.6 \times 10^{15} \, \mathrm{m}^2$. 
Simultaneously, the correction $\delta r$ already after one period 
reaches $1\%$ of the semi major axis, whose uncertainty can be measured 
with a much better accuracy of $4 \times 10^{-5} \, \%$. From this last 
value we incidentally find the rough limit 
$a \lesssim 1.7 \times 10^{14} \, \mathrm{m}^2$, if we cumulate 
$\delta r$ for one year.

The discovery of the double-pulsar binary PSR J0737-3039 paved the way to 
significantly improve the accuracy of binary pulsar measurements. An overview 
of the observed and derived parameters can be found in \cite{Bur,Lyn}.
With this data we evaluate the precession rate for the pulsar B 
predicted by GR to $5.07^{\circ}/\mathrm{yr}$. The measured value
$\Omega_{\textrm{B}} \approx 4.77^{+0.66}_{-0.65}\,{}^{\circ}/\mathrm{yr}$ 
\cite{BreKra} then allows the correction to lie within a minimal residue 
of $7 \, \%$ from $5.07^{\circ}/\mathrm{yr}$. This roughly yields the 
constraint $a \lesssim 2.3 \times 10^{15} \, \mathrm{m}^2$.

For other binary pulsars, even the required accuracy to precisely test GR  
is not yet reached by the corresponding experimental research, see e. g. 
\cite{ManKra} for the case of PSR J1141-6545. 
We finally conclude that a huge improvement of the accuracy of 
measurement would be necessary to put useful limits on $a$ by the 
precession of binary pulsars.

\section{Conclusions}

We gave the general formula for the lowest relativistic order coefficients of 
the $1/c$ expansion for the metric $g_{\mu\nu}$ of $f(R)$ gravity, where we considered 
functions of the form $f(R) = R + aR^2$. Furthermore, 
we investigated the GR and non-relativistic limits. The latter results in the Newtonian 
potential plus a Yukawa type correction with strength $G/3$ and Compton wave length 
$\sqrt{6a}$. As an application, 
we derived the $f(R)$-corrections to geodetic precession of orbiting gyroscopes.
The Lense-Thirring precession is not affected.

While the laboratory bound from the E\"ot-Wash experiment provides the small bound 
$a \lesssim 10^{-10} \, \mathrm{m}^2$, the results from Gravity Probe B imply the   
much larger limit $a \lesssim 5 \times 10^{11} \, \mathrm{m}^2$. The measurements 
of the precession of the pulsar B in the PSR J0737-3039 system provide instead the 
limit $a \lesssim 2.3 \times 10^{15} \, \mathrm{m}^2$.
Even for these large values of $a$ the quadratic term in (\ref{fR}) still 
induces a small correction of GR.

In principle, the coefficients for $g_{\mu\nu}$ can be used for the same 
applications as the PPN coefficients of metric gravity theories which have 
a Newtonian non-relativistic limit. However, the computation of the applications which 
require the fourth order coefficient ${}^{(4)}h_{00}$ are challenging, because 
its formula is quite involved and contains up to threefold iterated integrals. 
Therefore, a numerical analysis would be necessary for generic applications.

It would be interesting to take into account more general functions $f$. For 
instance one could extend the choice of $f$ to functions which are not necessarily 
analytic at $R=0$, but at a nonvanishing point $R=R_0$. Formally, this would imply 
to replace $f$ given in (\ref{fR}) by the more general function
\begin{equation}\label{fRg}
f(R) = -2\Lambda + a_1 R + a_2 R^2, \quad \Lambda,a_1,a_2 \neq 0.
\end{equation}
While the possibility of $a_1 \neq 1$ would not cause much trouble in the 
derivation of the $1/c$ expansion, the nonvanishing cosmological constant 
requires an expansion about a de Sitter or anti-de Sitter background, thus 
leading to more complicated partial differential equations for the potentials. 
Nevertheless, the choice of $f$ as in (\ref{fRg}) would be needed to study 
many $f(R)$ models which are proposed in the literature.

\begin{center}
{\bf Acknowledgements}
\end{center}

The authors wish to thank Micheal Kramer for useful discussions. Special thanks
go to Norbert Straumann for stimulating discussions and suggestions as well as for 
carefully cross-reading the article.

\end{document}